\documentstyle[12pt]{article}

\begin{document}
\def\la{\mathrel{\mathpalette\fun <}}
\def\ga{\mathrel{\mathpalette\fun >}}
\def\fun#1#2{\lower3.6pt\vbox{\baselineskip0pt\lineskip.9pt
  \ialign{$\mathsurround=0pt#1\hfill##\hfil$\crcr#2\crcr\sim\crcr}}}
\newcommand {\eegg}{e^+e^-\gamma\gamma~+\not \! \!{E}_T}
\newcommand {\mumugg}{\mu^+\mu^-\gamma\gamma~+\not \! \!{E}_T}
\renewcommand{\thefootnote}{\fnsymbol{footnote}}
\bibliographystyle{unsrt}

\begin{flushright} \small{
OSU-HEP-98-2\\
UMD-PP-98-98\\
March 1998}
\end{flushright}

\vspace{2mm}

\begin{center}
 
{\Large \bf A Four-Neutrino Mixing Scheme for Observed Neutrino
Data}\\

\vspace{1.3cm}
{\large S. C. Gibbons$^{1}$,   
R. N. Mohapatra$^{2}$, S. Nandi$^{1}$ and Amitava
Raychaudhuri$^{1,3}$\\
[5mm] 
$^1${\em Department of Physics, Oklahoma State University,\\
Stillwater, Oklahoma 74078, USA}\\
\vspace{2mm}
$^2${\em Department of Physics, University of Maryland, \\
College Park, Maryland 20742, USA}\\ 
\vspace{2mm}
$^3${\em Department of Physics, University of Calcutta,\\ 
92 Acharya Prafulla Chandra Road, Calcutta 700009, INDIA.}}
\end{center}

\vspace{2mm}
\begin{abstract}

It has been observed that simultaneous explanation of the solar
and atmospheric neutrino deficits and the reported evidence for
$\nu_{\mu}\rightarrow \nu_{e}$ oscillation from the Los Alamos
Liquid Scintillator Detector (LSND) requires at least one extra
neutrino species in addition to the three known ones. The extra
neutrino must be sterile with respect to the known weak
interactions. We present a new mass matrix for these four
neutrinos in which the LSND effect and the atmospheric neutrino
deficit are governed by only one parameter. We investigate the
phenomenological implications of such a mass matrix ansatz and
suggest possible ways to understand it in gauge theories.

\end{abstract}
 
\newpage
\renewcommand{\thefootnote}{\arabic{footnote})}
\setcounter{footnote}{0}
\addtocounter{page}{-1}
\baselineskip=24pt

\noindent{\bf I. Introduction}

These are very exciting times in neutrino physics. For the first
time in its history, there are several hints for a nonvanishing
mass for at least two of three known neutrinos which look very
promising and credible. They come (i) from the observation of
the solar neutrinos in various experiments and their disagreement
with the predictions of the standard solar model\cite{solar}: the
earlier experiments from Homestake, Kamiokande, SAGE and
GALLEX\cite{expt} and the most recent high statistics
confirmation of these results by the super-Kamiokande
experiment\cite{superK} and (ii) from the observation of the
atmospheric neutrinos by several previous
experiments\cite{atmos,fukuda} and the most recent confirmation
of the earlier results by the super-Kamiokande\cite{superK}
collaboration.  Then there is the result from the Los Alamos
Liquid Scintillator Neutrino Detector (LSND) which gives the
first laboratory evidence for the oscillation of both
$\bar{\nu}_{\mu}\rightarrow \bar{\nu}_e$\cite{LSND} as well as
$\nu_{\mu}\rightarrow \nu_e$ type\cite{LSND2}. The existence of
nonvanishing neutrino masses can be inferred from these data once
one tries to understand them in terms of neutrino oscillations
which depend on the mass difference square between the neutrino
species (i.e. $\delta m^2_{ij}$).

It has been pointed out that since the above processes 
require three different $\delta m^2$, it is impossible to
understand them within the conventional three neutrino
($\nu_e,~\nu_{\mu},~\nu_{\tau}$) framework; rather, one has to
include a sterile neutrino which is ultralight (denoted by
$\nu_s$)\cite{caldwell,carlo}. To see this in detail, note that
the MSW\cite{msw} type solution to the solar neutrino puzzle
including the latest super-Kamiokande results yields  the
following ranges (the so-called small angle solution) for the
neutrino mass differences and mixing\cite{solar1},
\begin{eqnarray} \delta m^2_{ei}\sim 4\times10^{-6} - 1.2 \times
10^{-5}{\rm eV}^2, \ \sin^22\theta_{ei}\sim 3 \times10^{-3} - 1.5
\times10^{-2}
\label{solar}
\end{eqnarray}
Next, a best fit to the presently available atmospheric neutrino
data from Kamiokande, IMB and Soudan as well as super-Kamiokande
assuming $\nu_{\mu}-\nu_{\tau}$ oscillation seems to
require\cite{gonz} at 90\% CL:
\begin{equation}
4\times 10^{-4}~{\rm eV}^2 \leq
\delta m^2_{atm} \leq 5 \times 10^{-3}~{\rm eV}^2, \;
\sin^22\theta_{atm}\simeq.76-1
\label{atm}
\end{equation}
Although the $\nu_e-\nu_{\mu}$
interpretation of this deficit for the older data from Kamiokande
is ruled out by the recent CHOOZ\cite{CHOOZ} results, inclusion
of the super-Kamiokande data still leaves an allowed domain for
$\delta m^2_{e\mu}$.  In this paper, we will assume
$\nu_{\mu}-\nu_{\tau}$ oscillation to be the explanation of the
atmospheric neutrino puzzle. Finally, the LSND data along with
the constraints from the reactor experiment at Bugey\cite{bugey}
and E776 at BNL\cite{e776} suggest that: 
\begin{equation}
.2~{\rm eV}^2\leq \delta
m^2_{e\mu}\leq 3~{\rm eV}^2, \;
2\times 10^{-3} \leq \sin^22\theta_{e\mu} \leq 4 \times 10^{-2}
\label{lsnd}
\end{equation}
We thus see that three different $\delta m^2$ values
are needed and hence the obvious need for an extra sterile
neutrino species.

Once one accepts the idea of four-way neutrino oscillation as a
necessity to solve the neutrino puzzles, the next question is
what is the nature of the mass matrix that can accommodate the
known data and finally what it implies for physics beyond the
standard model. It is the goal of this work to discuss the
first question followed by an attempt to study the resulting
theoretical implications.

\noindent{\bf II. The four neutrino mass matrix}

A simple mass matrix ansatz that embodies the above mixing
pattern was suggested in\cite{moh,barger} and analyzed in detail
in \cite{barger}.  This is given in the basis
$(\nu_s,\nu_e,\nu_{\mu},\nu_{\tau})$ by:
\begin{eqnarray}
M=\left(\begin{array}{cccc}
\mu_1 & \mu_3 & 0 & 0\\
\mu_3 & 0 & 0 & a\\
0 & 0 & \delta & m\\
0 & a & m & \delta \end{array} \right)
\label{theirm}
\end{eqnarray}
where one requires that $\mu_3\ll\mu_1, \delta\ll a \ll m$. A
characteristic of this mass matrix (to be called henceforth MI)
is that the atmospheric neutrino oscillation is governed by the
parameter $\delta$ whereas the mixing angle in the
$\nu_e-\nu_{\mu}$ oscillation is given by the parameter $a$. They
are unrelated.

Our proposal in this paper is to choose instead the following
mass matrix:
\begin{eqnarray}
M=\left(\begin{array}{cccc}
\mu_1 & \mu_3 & 0 & 0\\
\mu_3 & 0 & 0 & a \\
0 & 0 & \delta & m\\
0 & a & m & -\delta \end{array} \right)
\label{ourm}
\end{eqnarray}
This mass matrix is the same as the one in eq. (\ref{theirm})
except for the negative sign in the $\nu_{\tau}-\nu_{\tau}$
entry. We will call this MII. We will see that this simple change
of sign has profound phenomenological consequences. In particular,
it implies that the same parameter $a $ is responsible for
nonvanishing effects both for LSND as well as atmospheric
neutrinos.

In order to study the phenomenological implications of this mass
matrix, let us obtain the eigenvalues and the mixing matrix. We
will assume that the above mass matrices are defined in a basis
in which the charged leptons are mass eigenstates.

We choose the following hierarchy among the parameters:
$\mu_3\ll \mu_1 \ll a \ll \delta, m$. Note already a difference
from the mass texture MI that $\delta$ is much larger. To obtain
the two largest eigenvalues, we ignore the small parameters
$\mu_{1,3}$ and $a$.  It is then clear that the two eigenvalues
are equal in magnitude and opposite in sign i.e.
$\lambda_{3,4}=\pm \sqrt{m^2+\delta^2}$. This, of course, is not of
physical interest since it does not lead to atmospheric neutrino
oscillation. However, it gives us an idea of how big the
parameters $m$ and $\delta$ should be. For instance if
${\nu_{\mu,\tau}}$ are to constitute the hot dark matter of the
universe, then $\sqrt{m^2+\delta^2}$ should be in the eV range.

To get the mass splitting between the $\nu_{\mu}$ and
$\nu_{\tau}$ (mass eigenstates denoted by $\nu_3$ and $\nu_4$
respectively), we clearly have to keep the $a$ and consider the
$3\times 3$ matrix.  Diagonalizing it, we find that the masses of
$\nu_{\mu,\tau}$ are now split with
\begin{equation}
m_3^2 - m_4^2 \equiv 
\delta m^2_{atm}\simeq 
\frac{a^2\delta}{\sqrt{\delta^2+ m^2 + a^2}}
\label{matmos}
\end{equation}
Furthermore, the LSND $\nu_e-\nu_{\mu}$ mixing angle is given
(see eq. (\ref{probs})) by:
\begin{eqnarray}
\theta_{e\mu}\simeq \frac{a m}{m^2+\delta^2}
\label{emumix}
\end{eqnarray}
We thus see that in MII, the single parameter $a$ is responsible
for the atmospheric mass splitting as well as the LSND
$\nu_e- \nu_{\mu}$ mixing angle. This is different from what
happens in the MI case. We can therefore conclude that if
$\nu_e-\nu_{\mu}$ oscillation becomes very small and if the
$\delta m^2$ inferred from the atmospheric neutrino deficit
remains where it is, MII will be ruled out.  In fact, an even
stronger statement can be made i.e. if one takes the  $\delta
m^2$ values from the LSND experiment consistent with E776 and
Bugey (see eq. (\ref{lsnd})), one has $0.02
\leq \theta_{e\mu}\leq .1$. From eqs. (\ref{matmos}) and
(\ref{emumix}) $\delta m^2_{atm}
\simeq  \theta_{e\mu}^2 m\delta (\frac{m^2 +
\delta^2}{m^2})^{\frac{3}{2}}$. This would at most be $1.3\times
10^{-3}$ eV$^2$, implying that a significant part of the allowed
$\delta m^2_{atm}$ will be inconsistent with this model. In
other words, as the atmospheric neutrino data becomes more
precise MII will be testable unlike the MI texture where by
adjusting the different parameters $\delta$ and $a$ a much larger
range of values for the above experimental observables can be
accommodated.

Let us now turn to the $\nu_e-\nu_s$ mixing which 
is responsible for understanding the solar neutrino
puzzle in MII. For this purpose, we now consider the full $4\times 4$
mass matrix and find for the $\nu_e$ and $\nu_s$ (mass
eigenstates denoted by $\nu_2$ and $\nu_1$ respectively):
\begin{eqnarray} 
m^2_1-m^2_2\equiv \delta m^2_{solar}\simeq
\left(
\mu_1+\frac{a^2\delta}{m^2+\delta^2}\right)
\sqrt{\left(\mu_1-\frac{a^2\delta}{m^2+\delta^2}
\right)^2+4\mu^2_3}
\end{eqnarray}
To address the solar neutrino puzzle via the MSW mechanism it is
further necessary for the sterile neutrino to be heavier than the
$\nu_e$. 

In order to discuss further tests of the model, we write the
complete mixing matrix among the four neutrinos in the following
form:
\begin{eqnarray}
U=\left(\begin{array}{cccc}
1 & \epsilon_1 & 0 & 0\\
-\epsilon_1 & 1 & \epsilon_2 & \epsilon_3 \\
0 & \epsilon_4 & c & s \\
0 & \epsilon_5 & -s & c \end{array} \right)
\end{eqnarray}
where 
\begin{eqnarray}
s  \simeq -\sqrt{\frac{C-\delta}{2C}}; \;\;\;
c  \simeq \sqrt{\frac{C+\delta}{2C}};\;\;\; \nonumber \\
\epsilon_1 \simeq \frac{-\mu_3}{\mu_1-\frac{a^2\delta}{m^2+\delta^2}};
\;\;\;
\epsilon_2  \simeq \frac{a}{C}\sqrt{\frac{C-\delta}{2C}}
\nonumber \\
\epsilon_3  \simeq -\frac{a}{C}\sqrt{\frac{C+\delta}{2C}};\;\;\;
\epsilon_4  \simeq  -\frac{am}{C^2}; \;\;\;
\epsilon_5 \simeq  \frac{a\delta}{C^2} \nonumber\\
\label{elem}
\end{eqnarray}
and $C=\sqrt{m^2 + \delta^2} = \sqrt{\delta m^2_{e\mu}}$. Note
that $C$ and $c$ refer to different quantities.
 
In terms of these mixings, we can write down the oscillation
probabilities for various neutrino species $P(\nu_i\rightarrow
\nu_j)$ as follows:
\begin{eqnarray}
P(\nu_e\rightarrow \nu_e)& \simeq & 1 - 4 \epsilon_1^2 \sin^2
\Delta_s - 4(\epsilon_2^2 + \epsilon_3^2)\sin^2\Delta
\nonumber \\
P(\nu_{\mu}\rightarrow \nu_{\tau}) & \simeq & 4 c^2s^2 
\sin^2\Delta_{43}\nonumber\\
P(\nu_e\rightarrow \nu_{\mu})& \simeq & 4 cs\epsilon_2\epsilon_3
\sin^2\Delta_{43}-4\epsilon_4(c\epsilon_2+s\epsilon_3)\sin^2\Delta
\nonumber \\
P(\nu_e\rightarrow \nu_s) & \simeq & 4\epsilon^2_1
\sin^2\Delta_{21}\nonumber\\
P(\nu_e\rightarrow \nu_{\tau}) & \simeq & -4\left[
-\epsilon_2\epsilon_3 cs \sin^2\Delta_{43} +
\epsilon_5(-s\epsilon_2+c\epsilon_3)
\sin^2\Delta\right]\nonumber\\
P(\nu_{\mu}\rightarrow\nu_s) & = &0
\label{probs}
\end{eqnarray} 
where we have defined $\Delta_{ij}\equiv \frac{1.27\delta
m^2_{ij}L}{E}$ with four of the $\Delta_{ij}$ equal, i.e.
$\Delta_{42}=\Delta_{41}=\Delta_{32}=\Delta_{31}\equiv \Delta$.
The remaining two $\Delta_{ij}$ are very different and
explicitly included. We have used the usual units i.e. $\delta
m^2$ is in eV$^2$; $L$ is in kilometers and $E$ is in GeV.

The allowed ranges of the parameters where all constraints are
satisfied are presented in Table 1.
\parindent=0cm
\begin{center}
\begin{tabular}{|c||c|} \hline
Parameter & Allowed Range\\ \hline
$m$ & 0.39 eV -- 1.73 eV\\ \hline
$\delta$ & $2.2 \times 10^{-3}$ eV -- 0.85 eV \\ \hline
$a$ & $2.0 \times 10^{-2}$ eV -- 0.2 eV \\ \hline
$\mu_1$ & $5.5 \times 10^{-3}$ eV -- $6.5 \times 10^{-3} $eV \\ \hline
$\mu_3$ & $9.0 \times 10^{-6}$ eV -- $1.3 \times 10^{-4} $eV \\ \hline
\end{tabular}
\end{center}
{\bf Table 1:} The allowed ranges for the parameters of the model
consistent with present accelerator, reactor, solar and
atmospheric neutrino results.

\parindent=1cm
\vskip 20pt

These ranges must be taken together with the constraints:
\begin{equation}
m = C \sin(2\theta_{atm}); \;\;\;\;\;\; 
\delta = C \cos(2\theta_{atm}); \;\;\;\;\;\; 
\frac{a^2\delta}{C}  = \frac{\delta m^2_{atm}}{2}  
\end{equation}
\begin{equation}
3.3 \times 10^{-4} \leq \frac{\mu_1 C^2 - a^2 \delta}{C^2} \leq
2.1 \times 10^{-3}
\label{mu1}
\end{equation}
The ranges in Table 1 and the above constraints are based on the
assumption that $\mu_1$ is positive. If $\mu_1$ is negative then
(a) the constraint (\ref{mu1}) above does not apply, (b) Table 1
refers to the modulus of $\mu_1$, and (c) the range for $\mu_3$
changes to $1.5 \times 10^{-4} \leq \mu_3 \leq 7.4 \times
10^{-4}$.

To summarize this section, we note that in our mass matrix we use
five parameters to accommodate six experimental numbers i.e.
three mass difference squares and three mixing angles for the
solar, atmospheric and LSND experiments subject to the
constraints from the Bugey and E776 results. (We have checked
that the results from the CHOOZ and Fermilab E531\cite{E531}
experiments are automatically satisfied when the parameters are
in the allowed ranges.)  More importantly, because of the
intimate link between the $\theta_{e\mu}$ and $\delta m^2_{atm}$,
our mass texture MII is more easily testable than the MI texture.

\noindent{\bf III. Other experimental tests}

Turning now to the tests of the model, the first well known point
to emphasize is that the $\nu_e\rightarrow \nu_s$ oscillation
solution to the solar neutrino problem will be tested once the
SNO measures the neutral current effects of the solar neutrinos.
Unlike the $\nu_e\rightarrow
\nu_{\mu}$ oscillation scenario, where one would expect
$\Phi_{CC}/\Phi_{NC}\simeq .4$, for the $\nu_e\rightarrow \nu_s$
case, one should get $\Phi_{CC}/\Phi_{NC}\simeq 1$. Here $\Phi$
represents the signal for the appropriate process before the
incorporation of the cross-sections and detection efficiencies.

The second well known test of the model is of course in the long
baseline search for $\nu_{\mu}\rightarrow\nu_{\tau}$ oscillation
in experiments such as MINOS, where the oscillation probability
is driven by $\Delta_{atm}$ and  we expect
\begin{eqnarray}
\Delta_{atm}\simeq \left(\frac{\delta m^2_{atm}}{5\times
10^{-3}
{\rm eV}^2}\right)\left(\frac{L}{157~{\rm km}}\right)\left(\frac{\rm
GeV}{E}\right)
\end{eqnarray}
For $E=~10$ GeV, for the MINOS experiment this would imply a
transition probability of about 25\% for $\delta m^2\simeq
5\times 10^{-3}$ eV$^2$.

Let us now turn to $\nu_e\rightarrow \nu_{\tau}$ oscillation. In
this case, there is a difference between the MI and the MII
scenarios. In both scenarios there are two mass scales that
contribute to this probability: the one that governs the
atmospheric neutrino deficit and the other which is associated
with the LSND effect. However, from eqs.  (\ref{elem},
\ref{probs}) it is obvious that in the MII scenario the
contribution of the $\sin^2\Delta$ term is considerably larger
than in the MI case where $\delta \ll m$. Taking values for
$\epsilon_2\simeq
.02$, $\epsilon_3\simeq -.03$, and $\epsilon_5 \simeq .03$ as
dictated by our choice of parameters given earlier, we expect
that in a long baseline experiment,
\begin{eqnarray}
P(\nu_e\rightarrow \nu_{\tau})\simeq 3\times 10^{-4}+ 10^{-3} 
\sin^2\Delta_{atm}
\end{eqnarray}
In the MI scenario, the first term is negligible. 

The predicted $\nu_e \rightarrow \nu_{\tau}$ oscillations driven by two
vastly different mass-scales but of comparable strength could be
observable in future facilities (presently under contemplation)
such as the muon collider with a long baseline neutrino
experiment\cite{geer}. For the purpose of illustration we
consider the detectability of a $\nu_{\tau} $ signal through the
$\tau^- \rightarrow \mu^-$ decay where the parent $\nu_e$ beam of
$E_{max}$ = 20 GeV originates from the decay of $\mu^+$s produced
for a muon collider set-up at Fermilab. Our results are presented
in the Figure. We find that a 10 kT detector at Soudan will be
able to scan, at the single event/year signal level, significant
portions of the predicted mixing angle -- mass-square-difference
space of both the mass scales but Gran Sasso will be
sensitive to a smaller part of only the LSND mass-scale. If
a 10 kT detector could be set-up at a distance of 10 km from the
source then its reach in the mixing angle will be much wider and
we show in the figure the sensitivity of such a facility assuming
100 signal events/year.

There will also be testable difference between the MI and the MII
scenarios in the short baseline experiments. In this case, the
$\Delta_{34}$ contribution to $P(\nu_e\rightarrow \nu_{\tau})$
would be insignifcant. As a result, in the MI scenario, one would
expect this oscillation to be negligible. On the other hand in
the MII scenario, we get
\begin{eqnarray}
P(\nu_e\rightarrow \nu_{\tau})\simeq 10^{-3} \sin^2\Delta
\end{eqnarray}
This is large enough to be observable with a reasonable $\tau$
detection efficiency.

In our model, the neutrinoless double beta decay 
vanishes at the tree level since its amplitude is proportional to 
$<m_{\nu_e}>$ which is given by 
\begin{eqnarray}
<m_{\nu_e}>=\Sigma_i U^2_{ei} \eta_i m_{\nu_i}=M_{\nu_e\nu_e}
\end{eqnarray}
Here $M_{\nu_e\nu_e}$ is the $\nu_e-\nu_e$ element of the
neutrino mass matrix in the weak basis and $\eta_i = \pm 1$ is
the sign of the $i$-th eigenvalue. As is clear from eq.
(\ref{ourm}), this entry vanishes in our case. 
This also happens in the scenario MI,
contrary to the assertion in Ref. \cite{barger}.

\vspace{4mm}

\newpage

\noindent{\bf IV. Possible theoretical scenarios}

In this section, we explore possible theoretical scenarios that
could lead to the mass texture analyzed in this paper.
Specifically, we want to reproduce the key distinguishing feature
of the model which is
$M_{\nu_{\mu}\nu_{\mu}}=-M_{\nu_{\tau}\nu_{\tau}}$ within a
reasonable scenario that can also roughly account for the rest of
the texture.

Let us consider an extension of the standard model where the
fermion sector is left untouched but the Higgs sector has several
additional scalars and the theory is invariant under the extra
global symmetry $S_3\times Z_2$ where $S_3$ is the permutation
symmetry on three objects. The extra Higgs fields included are
two triplets ($\Delta_{i}$) with lepton number $L=-2$, which
acquire a tiny vev due to a Higgs see-saw mechanism; four
doublets $H_{1,2}$ and $H_{3,4}$ with identical quantum numbers to the
standard model doublet and different quantum numbers under the discrete
global symmetry group; two iso-singlet 
charged ($\eta^+_{i}$) fields with $L=+2$; a doubly charged iso-singlet
field ($k^{++}$) with $L=-2$ and a neutral $L=2$ field $\sigma$.
We will assume that $<\sigma >\gg <H^0_{2,3,4}>
\gg < \Delta^0_{i}>\simeq $ 1-2 eV. We further assume that $<H^0_1>=0$.
The $S_3\times Z_2$ transformation property of the Higgs and
fermion fields are presented in Table 2.

\parindent=0cm

\begin{center}
\begin{tabular}{|c|c|c|c|c|c|c|c|c|c|c|} \hline
Fields & $\pmatrix{L_{\mu}\cr L_{\tau}\cr}$ & $L_e$ &
$\pmatrix{\mu^c \cr \tau^c \cr}$ & $e^c$ &$H_{3,4}$ & $\pmatrix{H_1
\cr H_2 \cr}$ &  $\pmatrix{\Delta_1 \cr \Delta_2 \cr}$ & $k^{++}$ & 
$\pmatrix{\eta_1 \cr \eta_2 \cr}$ & 
$\sigma $  \\  \hline
$S_3\times Z_2$& & & & & & & & & & \\
quantum & (2,+) & (1,-) & (2,+)& (1,-) & (1, +,-) & (2, +) &
(2,+)  & (1,+) & (2, -) & (1,+) \\ 
numbers & & & & & & & & & & \\ \hline
\end{tabular}
\end{center}
{\bf Table 2:} The lepton and scalar fields of the model and their
$S_3\times Z_2$ quantum numbers. There are two one-dimensional
representations of $S_3$. Here 1 stands for the invariant
representation.  

\vskip 20pt
\parindent=1cm

The Yukawa Lagrangian invariant under the gauge group and the
extra global symmetries is:
\begin{eqnarray}
{\cal L} & = &  f_1\left[2 L_{\tau}L_{\mu}\Delta_1+(L_{\mu}L_{\mu}- 
L_{\tau}L_{\tau}) \Delta_2 \right]\nonumber\\
& & +h_1 L_eH_3 e^c +h_2(L_{\mu}\mu^c+L_{\tau}\tau^c)H_3\nonumber\\
& & +h_3 \left[ (L_{\mu}\tau^c+L_{\tau}\mu^c)H_1 + (L_{\mu}\mu^c 
-L_{\tau}\tau^c) H_2\right]\nonumber\\
& & +f_2 (L_{\tau}\eta_2 +L_{\mu}\eta_1)L_e +f_3 (\tau^c\tau^c 
+\mu^c\mu^c)k^{--} +f_4 e^ce^ck^{--}
\end{eqnarray}
The Higgs Lagrangian will contain two parts: one part consisting
only of dimension four operators that are invariant under the
gauge as well as the global symmetry and a part consisting of
terms with dimension two or three which will softly break the
global symmetries. In the equation below, we give only the part
of the potential relevant to generating the neutrino Majorana
masses out of two loop effects.
\begin{eqnarray}
{\cal L}_{Higgs} = V_0 + \mu_1 ({\eta^{+}_1}^2+{\eta^{+}_2}^2)k^{--} +
\lambda (\Delta_1H_1
+\Delta_2H_2)H_3\sigma + m^2_0 \Delta^{\dagger}_1\Delta_2\nonumber \\
+\lambda'H_4[(\Delta_1H_2+\Delta_2H_1)\eta_1+(\Delta_1H_1-\Delta_2H_2)\eta_2]
+h.c.
\end{eqnarray}

Let us now study the tree level mass matrix for the three known
neutrinos in this model. It is given by
\begin{eqnarray}
M_0=\left(\begin{array}{ccc}
0 & 0 & 0\\
0 & f_1v_2 & f_1v_1\\
0 & f_1v_1 & -f_1v_2 \end{array}\right)
\end{eqnarray}
where $v_i=<\Delta^0_i>$. Thus this has the desired structure for
the scenario MII in the $\nu_{\mu}-\nu_{\tau}$ sector.
Incidentally, the $\lambda$ term in the Higgs Lagrangian can be
used to suppress the $\Delta_i$ vev's via a see-saw mechanism.

The $\nu_e-\nu_{\tau}$ entry (i.e. $a$) in
the mass matrix arises via a two loop diagram. The magnitude
for this term can be estimated to be
\begin{eqnarray}
a \simeq \frac{f^3\mu^3m^2_{\tau}}{(16\pi^2)^2M^4}
\end{eqnarray}
where $M$ is a typical scalar mass.
If we choose $f_i\approx 4\times 10^{-3}$ and $\mu\approx 10 M\simeq 1$ TeV, 
then 
we get $a\approx .01$ eV which is of the right order of magnitude. 
We have checked that the contributions to other entries of the neutrino
mass matrix are adequately small. It is
important to remark that for our mixing considerations to hold,
the charged leptons must be diagonal. This is guaranteed by
$<H_1>=0$. We have to fine tune $h_2<H_3>$
and $h_3<H_2>$ to get the muon and the tau lepton masses right.

So far we have ignored the sterile neutrino. It can be
incorporated into the model in one of two ways. An elegant way to
do this is to double the particle spectrum and the gauge forces
of the standard model by including a mirror sector to the
standard model as is done in Ref.\cite{bere} and inducing the
Majorana masses for the neutrinos via non-renormalizable
operators. The lightest neutrino of the mirror sector will be
identified with the sterile neutrino.  The non-renormalizable
operators will in general induce nonzero values in the
$\nu_e-\nu_{\mu}$ entry as well as in the $\nu_e-\nu_e$ entry but
these contributions are in the range of $10^{-6}$ eV so that they
do not affect any of our considerations. The mirror
$\nu'_{\mu,\tau}$ masses will arise from triplets whose vevs will
be larger than the corresponding ones in the visible sector since
the see-saw mechanism giving their vevs is proportional to
doublet vevs, which for cosmological reasons must be larger  in
the mirror sector: $<H_{mirror}>\simeq 30 <H>$.

Another way is to add the sterile neutrino as an extra fermion to
the standard model and use the radiative mechanism to suppress
its mass to the desired level. By adding new Higgs fields to the
model already described, this can be accomplished. We do not
elaborate on this here.

\vspace{4mm}

\noindent{\bf V. Conclusion}

In conclusion, we have presented a new four neutrino mass matrix
which fits all neutrino observations with five parameters and
where the $\nu_e-\nu_{\mu}$ mixing and the $\delta m^2_{\mu
\tau}$ are related to each other unlike the simplest four
neutrino mass matrix discussed in the literature. Thus, in this
scenario the atmospheric neutrino deficit seen in the present
data implies the observable effect in  $\nu_{\mu}-\nu_e$
oscillation as found at LSND. This model can also be tested in the
short as well as long baseline $\nu_e-\nu_{\tau}$ oscillation
searches. We also construct a theoretical scenario which could
lead to the mass matrix of the type under consideration.

 \vspace{4mm}
{\bf Acknowledgement}
\vspace{4mm}

The work of S. C. G., S. N., and A. R. is supported by the U.S.
Department of Energy while that of R. N. M. is supported by a
grant from the National Science Foundation, grant no.PHY-9421385.
A. R. is also partially supported by grants from C.S.I.R., India 
and D.S.T., India. He is grateful to the Department of Physics of the
Oklahoma State University for their hospitality.


\newpage
\begin{center}
{\bf FIGURE CAPTION}
\end{center}
The prediction of this model for $\nu_e - \nu_{\tau}$ oscillation
in an effective two-flavor plot is shown as the rectangular
regions. The sensitivity of experiments located at Gran Sasso
(L=9900 km), Soudan (L=732 km) and a new facility (L=10 km) are
shown when the source is a $\nu_e$ beam obtained from a parent
muon beam (see text) at Fermilab with
$E_{max}$ = 20 GeV. The Gran Sasso and Soudan results correspond to
a single-event signal/year while for the new facility at 10 km the
results shown correspond to 100 signal events/year.
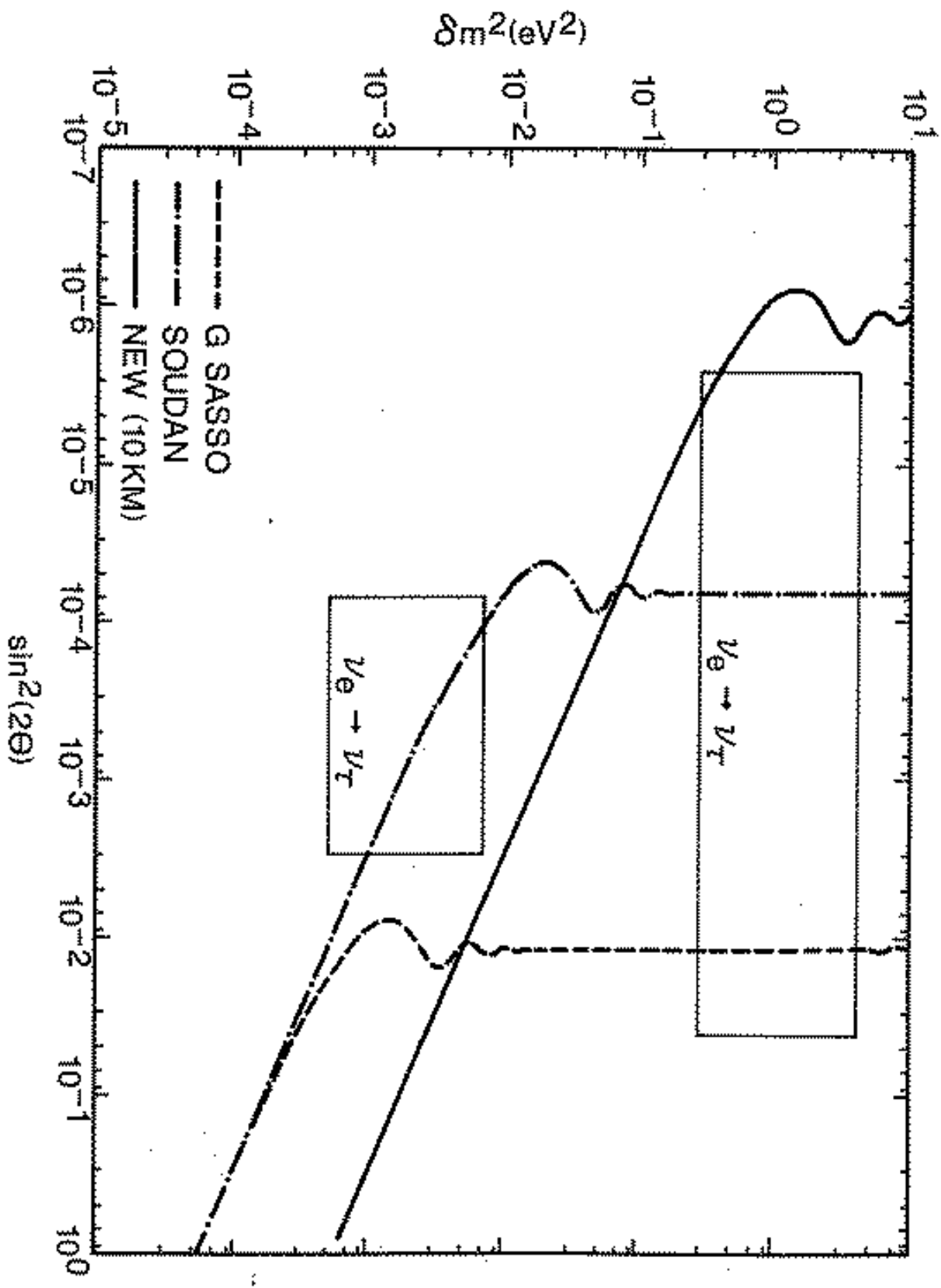
\end{document}